\documentclass[aps,showpacs,prd,12pt]{revtex4}

\begin{document}

\def\be{\begin{equation}}
\def\ee{\end{equation}}
\def\bea{\begin{eqnarray}}
\def\eea{\end{eqnarray}}

\title{Temperature inversion symmetry in the Casimir 
effect with an antiperiodic boundary condition}

\author{A.C. Aguiar Pinto}\email{acap@if.ufrj.br}
\author{T.M. Britto}\email{britto@if.ufrj.br}  
\author{F. Pascoal}\email{fabiopr@if.ufrj.br}
\author{F.S.S. da Rosa}\email{siqueira@if.ufrj.br}
\affiliation{Universidade Federal do Rio de Janeiro, Instituto de F\'{\i}sica\\
Caixa Postal 68528, CEP.: 21945-970, Rio de Janeiro-RJ, Brazil.}
\date{\today}
\begin{abstract}
We present explicitly  another example of a temperature inversion symmetry in the Casimir 
effect for a nonsymmetric boundary condition. We also give an interpretation for our result.
\end{abstract}
\pacs{11.10.Wx, 12.20.Ds, 33.15.-e}
\maketitle                 
%
%
This brief report was motivated by a recent paper published 
by Santos {\it et al.} \cite{Fila}, in which they discuss the 
temperature inversion symmetry in the Casimir effect \cite{casimir1948} 
for mixed boundary conditions (for a detailed discussion on the Casimir effect see 
\cite{PlunienGreinerMuller86,BordagMostepanenkoMohideen2001} and references therein). In an earlier paper,  
Ravndal and Tollefsen \cite{ravndal1989} showed that for the 
usual setup  of two parallel plates  a simple inversion symmetry 
arises in the Casimir effect at finite temperature.
Temperature inversion symmetry also 
appeared in the Brown-Maclay work \cite{Brown1969} where they 
related directly the zero-temperature Casimir energy to the 
energy density 
of blackbody radiation at temperature $T$. A few other papers on 
this kind of symmetry have also been published \cite{tadaki86, 
gundersen1988, lutken1988, wotzasek1990, fila2000}.
Until the publication of Ref.\cite{Fila}, 
this kind of inversion symmetry had appeared only in 
calculations of Casimir energy involving 
symmetric boundary conditions. In 1999, Santos {\it et al.} \cite{Fila} 
showed, for the case of a massless scalar field submitted to 
mixed boundary conditions (Dirichlet-Neumann), 
 that the Helmholtz free energy per unit area could be 
written as a sum of two terms, each of them obeying separately 
a temperature inversion symmetry. Our purpose here is to present 
another kind of nonsymmetric boundary 
condition for which there exists such a symmetry. We show 
explicitly that for the massless 
scalar field under an antiperiodic boundary condition the 
Helmholtz free energy per unit area can also be cast 
as a sum of two terms, where each one 
satisfies a temperature inversion symmetry.

The Casimir effect for a massless scalar field under an 
antiperiodic boundary condition (compactification of R$^1$ to S$^1$) 
in a $3+1$ spacetime is given by
\begin{equation}
\phi(\tau,x,y,z) =-\phi(\tau,x,y,z+a). \label{cond_anti}
\end{equation}
  
We will use the imaginary time formalism, which means that 
\begin{equation}
\phi(\tau,x,y,z) = \phi(\tau+\beta,x,y,z) \label{cond_per}
\end{equation}
where $\beta = T^{-1}$, the reciprocal of the 
temperature. The conditions (\ref{cond_anti}) and (\ref{cond_per}) 
lead us to the following eigenvalues for the euclidean 
operator $\partial_E^2=\partial^2 /\partial \tau^2 + \nabla^{2}$:
\begin{equation}
\left\{\kappa^2+ (2n+1)^2 \left(\frac{\pi}{a}\right)^2 + 
\left(\frac{2\pi\ell}{\beta}\right)^2 
\mbox{where}\; \kappa^2=k_1^2+ k_2^2\; ; \;\;  n \;\; 
{\rm and} \;\; 
\ell = 0, \pm 1, \pm 2, \cdots \right\}. \label{autovalores}
\end{equation}

The partition function for a massless scalar 
field at finite temperature is given by:
\begin{equation}
{\cal Z}(\beta) = N \int_{periodic} [{\cal D}\phi] \; exp{\int_0^\beta 
\int d^3 x {\cal L}},
\end{equation}
where $N$ is a normalization constant and ${\cal L}$ is the 
Lagrangian density for the theory under study. The Helmholtz free 
energy $F(\beta)$ can then be written in terms of the 
corresponding generalized $\zeta$ function as
\begin{equation}
F(\beta) = - \frac{1}{2\beta} \frac{d}{ds} 
\zeta (s,-\partial_E^2){\Big |}_{s=0} \label{helmholtz}\; .
\end{equation}
After some algebra, we can write the zeta function as
\begin{eqnarray}
\zeta (s,-\partial_E^2)&=& \frac{L^2}{4\pi} \frac{\Gamma(s-1)}{\Gamma(s)} 
\Big\{ 2 \Big(\frac{\pi}{a}\Big)^{2-2s} (1- 2^{2-2s})\zeta_R(2s-2) + 
\nonumber \\
\nonumber \\
&+& 4 \pi^{2-2s} E_2\left(s-1,\frac{1}{a^2}, \frac{4}{\beta^2}\right)- 
4 \pi^{2-2s} E_2\left(s-1,\frac{4}{a^2}, \frac{4}{\beta^2}\right)\Big\}\; ,
\label{zeta2}
\end{eqnarray}
where $\zeta_R(z)$ is the Riemann zeta function and 
$E_2(z,a_1,a_2)$ is an Epstein function. Using the 
same methods of Ref.\cite{Fila}, 
we can write the Helmholtz free energy per unit area as:
\begin{equation}
\frac{F(\beta)}{L^2} = \frac{7}{720}\frac{\pi}{a^3} - 
\frac{1}{\pi\beta^3} f(\xi), \label{F2}
\end{equation}
where we defined $\xi= a/\pi\beta$ and
\begin{equation}
f(\xi) = \frac{1}{2\pi^4\xi^3} \Big\{\sum_{\ell,n=-\infty}^\infty 
\frac{(-1)^n \pi^4\xi^4}{[\ell^2+\pi^2\xi^2 n^2]^2}
- \sum_{n=-\infty}^\infty\frac{(-1)^n}{n^4}\Big\}.\label{def_f_Xi}
\end{equation}
Note that the first term on the r.h.s. of Eq.(\ref{F2}) represents the 
Casimir energy at zero temperature for the antiperiodic 
boundary condition.
We should now eliminate from Eq.(\ref{def_f_Xi}) the terms 
$\ell=n=0$ in the 
first summation and $n=0$ in the second one (see Ref.\cite{Fila} 
for details). 
Once it has been done, we have (after convenient manipulations): 
\begin{equation}
\frac{F(\beta)}{L^2} = \frac{F_1(\beta)}{L^2} - \frac{F_2(\beta)}{L^2}\; , 
\label{F4}
\end{equation}
where the functions $\frac{F_1(\beta)}{L^2}$ and 
$\frac{F_2(\beta)}{L^2}$ are defined by
\begin{equation}\label{defF1}
\frac{F_1(\beta)}{L^2} = -\frac{1}{16\pi^2 a^3} \sum_{\ell,n=-\infty}^\infty 
\frac{(2\pi\xi)^4}{[\ell^2+(2\pi\xi n)^2]^2} 
\end{equation}
and
\begin{equation}\label{defF2}
\frac{F_2(\beta)}{L^2} = -\frac{1}{2\pi^2 a^3} \sum_{\ell,n=-\infty}^\infty 
\frac{(\pi\xi)^4}{[\ell^2+(\pi\xi n)^2]^2}\; ,
\end{equation}
which satisfy the following temperature inversion symmetry relations:
\begin{equation}
F_1(\xi) = (2\pi\xi)^4 F_1\Big( \frac{1}{4\pi^2 \xi} \Big) 
\;\;\;\;\; \mbox{and} \;\;\;\;\;  
F_2(\xi) = (\pi\xi)^4 F_1\Big( \frac{1}{\pi^2 \xi} \Big). 
\label{def_F4}
\end{equation}
The tempereture inversion symmetry just presented for the case 
with an antiperiodic boundary condition may be interpreted 
following the same lines as that appearing in Ref.\cite{Fila}. 
 In this reference, Santos {\it et al.} showed that the 
Helmholtz free energy per unit 
area for a massless scalar field under mixed boundary 
condition may be written 
as a sum of two terms corresponding, each one, to a pair 
of  uncharged parallel 
perfectly conducting plates kept at a distance $2d$ and $d$ apart, 
respectively. 
In the present case, we have an analogous situation, namely, the 
Helmholtz free energy per unit area for a massless scalar field 
under an antiperiodic boundary condition (with spatial \lq\lq 
period{\rq\rq} $a$) may also be written 
as a sum of two terms corresponding, each one, to a periodic 
boundary condition, but with 
spatial periods $2a$ and $a$, respectively (see Eqs.(\ref{defF1}) 
and (\ref{defF2})). 

Temperature inversion symmetry directly relates the Casimir effect at 
zero temperature to its high temperature limit, where the Stefan-Boltzmann term 
dominates and hence may be viewed as one of the simplest examples of duality. Our 
result provides one more explicit example of such a phenomenon. In addition, this 
kind of symmetry can be useful to derive approximate expressions for the Helmholtz 
free energy. For instance, we can derive the low temperature limit, that is, we may calculate  
the first thermal corrections to the zero temperature Casimir energy, from the 
high temperature limit, which is in general much easier to obtain. 

The consideration of massive fields is not an easy task. It is not obvious  
whether this duality symmetry will remain valid for massive fields, but this will 
be left for a future investigation.


\end{document}